\documentclass[showpacs,titlepage,prd,aps,
preprint,
amsmath,byrevtex,nofootinbib]{revtex4}
\usepackage{graphicx}     
\usepackage[dvips]{epsfig}
\newcommand{\bea}{\begin{eqnarray}}   
\newcommand{\eea}{\end{eqnarray}}

\newcommand{\deltaunodue}{\mbox{$\Delta m_{21}^2 $}}
\newcommand{\deltaunotre}{\mbox{$\Delta m_{31}^2 $}}

\newcommand{\nova}{\mbox{NO$\nu$A} }

\newcommand{\beq}{\begin{eqnarray}}
\newcommand{\eeq}{\end{eqnarray}}

\begin{document}

\begin{flushright}
FERMILAB-PUB-07-456-APC-E-T\\[-0.1in]
ROMA-TH-1458\\[-0.1in]
\end{flushright}


\title{Combining CPT-conjugate Neutrino channels at Fermilab}
\author{
Andreas Jansson$^1$,
Olga Mena$^2$,
Stephen Parke$^1$ and 
Niki Saoulidou$^1$}
\address{$^1$
\sl Fermi National Accelerator Laboratory \\
P.O.Box 500, Batavia, IL 60510, USA}
\address{
$^2$
\sl INFN Sez.\ di Roma,
Dipartimento di Fisica, Universit\`{a} di Roma``La Sapienza'',\\
P.le A.~Moro, 5, I-00185 Roma, Italy}
 
\date{\today}

\vglue 1.4cm

\begin{abstract}
We explore an alternative strategy to determine the neutrino mass hierarchy by making use of possible future neutrino facilities at Fermilab. Here, we use CPT-conjugate neutrino channels, exploiting a $\nu_\mu$ beam from the NuMI beamline and a $\bar{\nu}_e$ beam from a betabeam experimental setup.
Both experiments are performed at approximately the same $\langle E \rangle /L$. We present different possible accelerator scenarios for the betabeam neutrino setup and fluxes. 
This CPT-conjugate neutrino channel scenario can extract the neutrino mass hierarchy down to
$\sin^2 2 \theta_{13} \approx 0.02$.
\end{abstract}
\maketitle
\section{Introduction}
During the last several years the physics of neutrinos has achieved
remarkable progress. The experiments with solar~\cite{sol,SKsolar,SNO1,SNO2,SNO3,SNOsalt},  atmospheric~\cite{SKatm}, reactor~\cite{KamLAND}, and also
long-baseline accelerator~\cite{K2K,MINOSprop,MINOS} neutrinos,
have provided compelling evidence for the existence of neutrino oscillations, implying non zero neutrino
masses. The present data require two large
($\theta_{12}$ and $\theta_{23}$) and one small ($\theta_{13}$) angles
in the neutrino mixing matrix~\cite{BPont57}, and at least two mass squared differences,
$\Delta m_{ji}^{2} \equiv m_j^2 -m_i^2$ (where $m_{j}$'s are the neutrino
masses), one driving the atmospheric ($\deltaunotre$) and the other one the solar ($\deltaunodue$) neutrino oscillations. The mixing
angles $\theta_{12}$ and $\theta_{23}$ control the solar and the
dominant atmospheric neutrino oscillations, while $\theta_{13}$ is the
angle limited by the data from the CHOOZ and Palo Verde reactor
experiments~\cite{CHOOZ,PaloV}.

The Super-Kamiokande (SK)~\cite{SKatm} and K2K~\cite{K2K} data are well
described in terms of dominant $\nu_{\mu} \rightarrow \nu_{\tau}$
($\bar{\nu}_{\mu} \rightarrow \bar{\nu}_{\tau}$) vacuum
oscillations. 
A recent global fit~\cite{concha} provides the following $3 \sigma$ allowed ranges for the atmospheric mixing parameters
\beq 
\label{eq:range}|\deltaunotre| =(2 - 3.2)\times10^{-3}{\rm eV^2},~~~~
0.32<\sin^2\theta_{23}<0.64~.
\eeq
The sign of $\deltaunotre$, sign$(\deltaunotre)$, 
cannot be determined with the existing data. The two possibilities,
$\deltaunotre > 0$ or $\deltaunotre < 0$, correspond to two different
types of neutrino mass ordering: normal hierarchy and inverted hierarchy.
In addition, information on the octant in which $\theta_{23}$ lies, if $\sin^22\theta_{23} \neq 1$, is beyond the reach of present experiments. 

The 2-neutrino oscillation analysis of the solar neutrino data,
including the results from the complete salt phase of the Sudbury
Neutrino Observatory (SNO) experiment~\cite{SNOsalt}, in combination
with the KamLAND spectrum data~\cite{KL766}, shows that the solar neutrino oscillation parameters lie in the low-LMA (Large Mixing Angle) region, with best fit values~\cite{concha} $\deltaunodue =7.9 \times 10^{-5}~{\rm eV^2}$ and $\sin^2 \theta_{12} =0.30$.


A combined 3-neutrino oscillation analysis of the solar, atmospheric,
reactor and long-baseline neutrino data~\cite{concha} constrains the third mixing angle to be $\sin^2\theta_{13} < 0.04$ at the $3\sigma$ C.L. However, the bound on $\sin^2 \theta_{13}$ is dependent on the precise value of $\Delta m^2_{31}$.

The future goals for the study of neutrino properties is to
precisely determine the already measured oscillation parameters
and to obtain information on the unknown ones: namely $\theta_{13}$,
the CP--violating phase $\delta$ and the type of neutrino mass
hierarchy (or equivalently sign$(\deltaunotre)$).  In the presence of matter effects, the neutrino (antineutrino) oscillation probability gets enhanced~\cite{matter,matterosc} for the normal (inverted) hierarchy. 
Making use of the different
matter effects for neutrinos and antineutrinos seems, in principle, the most promising way to distinguish among the two possibilities: normal versus inverted hierarchy. However, the sensitivity
to the mass hierarchy determination from the neutrino-antineutrino comparison is highly dependent on the value of the CP violating phase. Thus, possible alternative methods were first proposed in Ref.~\cite{MNP03}. 
In this paper we concentrate on the extraction of the neutrino mass hierarchy 
by combining a $\nu_\mu\to \nu_e$ experiment with its CPT conjugated channel 
$\bar{\nu}_e \to \bar{\nu}_\mu$, see Ref.~\cite{MNP03}. 
More recently, it is primarily the CPT-conjugate channel pairs that give the CERN-MEMPHYS proposal sensitivity to the hierarchy, see Ref.~\cite{memphys}. 
If nature respects CPT symmetry, then, at the same $E/L$ the only difference between the two flavor transitions can come from matter effects and that near the first oscillation maximum
\begin{eqnarray}
P(\nu_\mu \to \nu_e) & > & P(\bar{\nu}_e \to \bar{\nu}_\mu) \quad {\rm for  ~Normal ~Hierarchy}
\nonumber \\
{\rm and} \quad 
P(\nu_\mu \to \nu_e) & < & P(\bar{\nu}_e \to \bar{\nu}_\mu)  \quad {\rm for ~Inverted~Hierarchy,}
\nonumber
\end{eqnarray}
i.e. for the normal hierarchy the neutrino channel is enhanced and the antineutrino CPT conjugate
channel suppressed and vice versa for the inverted hierarchy.
This is the effect that will be exploited in this paper to determine the neutrino mass hierarchy.

We will show that the combination of the Phase I (neutrino-data only) of the long-baseline $\nu_e$ appearance experiment NO$\nu$A~\cite{newNOvA}, exploiting the off-axis technique\footnote{A neutrino beam with narrow energy spectrum can be produced by placing the detector off-axis, i.~e., 
at some angle with respect to the forward direction. The resulting neutrino spectrum is very narrow in energy 
(nearly monochromatic, $\Delta E /E \sim 15 - 25\%$) and peaked at lower energies with respect to the on-axis one.
The off-axis technique allows a discrimination between the peaked $\nu_e$ oscillation signal and the intrinsic 
$\nu_e$ background which has a broad energy spectrum~\cite{adamoff}.In addition, the off-axis technique
reduces significantly the background resulting from neutral current interactions of higher energy neutrinos with a $\pi^{0}$ in the final state.} with a possible future betabeam facility \cite{zucchelli,mauro,betabeampilar,betabeams,iss} at Fermilab exploiting a $\bar{\nu}_e$ neutrino beam from radiative ion decays could help enormously in measuring the neutrino mass hierarchy. 
For our analysis, unless otherwise stated, we will use a representative value of $|\deltaunotre| = 2.5 \times 10^{-3} \
\rm{eV}^2$ and $\sin^2 2 \theta_{23}=1$. 
For the solar oscillation parameters $\deltaunodue$ and $\theta_{12}$, we will use the best fit values
quoted earlier in this section. 
The structure of the paper is as follows. In Section~\ref{sec:strat} we present the general physics strategy used to determine  the neutrino mass hierarchy including the CPT conjugate channels used in this paper. Section~\ref{sec:beta} contains a realistic description of possible future betabeam facilities at Fermilab. The different scenarios deal with different ions, baselines and luminosities, and the performance of the strategy followed here in each of these scenarios is illustrated in Section~\ref{sec:possible}.  
 The sensitivity curves for the several scenarios will be presented in Section~\ref{sec:results} and the final remarks are summarized in Section~\ref{sec:conclusions}.
 In the Appendix~\ref{sec:cptintro}, we discuss the details associated with comparing
 CPT conjugate neutrino oscillation probabilities.

\section{Combining Neutrino channels}
\label{sec:strat}

The strategy we have introduced in the previous section and we explain in detail here is different from the usual one, which exploits the
combination of the neutrino and antineutrino oscillation channels. Typically, the proposed long baseline neutrino oscillation experiments
have a single far detector and plan to run with the beam in two different modes, muon neutrinos and muon antineutrinos. In principle, by
measuring the probability of neutrino and antineutrino flavor conversion, the values of the CP--violating phase $\delta$ and the
sign$(\deltaunotre)$ could be extracted, since, in the presence of matter effects there will be two allowed regions for each type of
hierarchy, normal or inverted, in the $P(\nu_ \mu \to \nu_e)$ versus $P(\bar \nu_\mu \to \bar \nu_e)$ plane. In practice, the neutrino--antineutrino comparison does not provide the ideal tool to extract the neutrino mas hierarchy, as we explain below.

\begin{figure}[t]
\begin{center}
\includegraphics[width=3in]{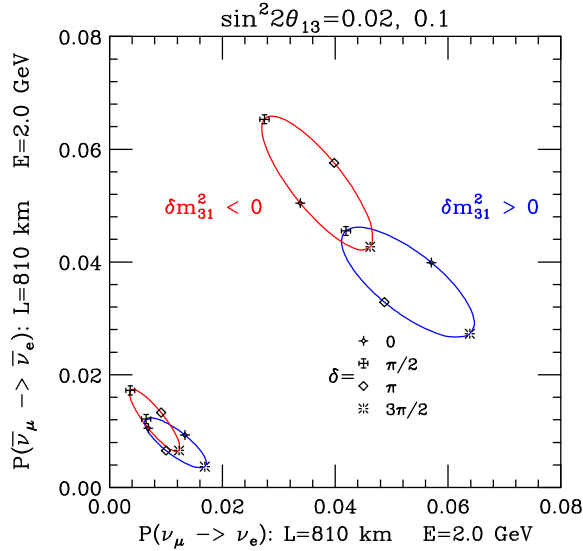}
\end{center}
\caption[]{\textit{The bi--probability plot for $P(\nu_\mu \to \nu_e)$ versus $P(\bar{\nu}_\mu \to \bar{\nu}_e)$ at a baseline of 810 km and an energy of 2.0 GeV for the normal (blue) and the inverted (red) hierarchies. The smaller, lower (larger, upper) ellipses are for $\sin^2 2 \theta_{13}=0.02$ ($~0.10$).}}
\label{fig:comp}
\end{figure}

Suppose we compute the oscillation probabilities $P(\nu_ \mu \to \nu_e)$ and $P(\bar \nu_\mu \to \bar \nu_e)$ for a given set of oscillation parameters and the CP-phase $\delta$ is varied between $0$ and $2 \pi$: we obtain a closed CP trajectory (an ellipse) in the bi--probability space of neutrino and antineutrino conversion~\cite{MN01}. Matter effects are responsible for the departure of the center of the ellipses from the diagonal 
line in the bi--probability plane for normal and inverted hierarchy. In Figure~\ref{fig:comp}, we have illustrated the case for $E=2.0$~GeV
and $L=810$~km, which roughly correspond to those of the NO$\nu$A experiment~\cite{newNOvA}. The distance between the center of the ellipse for the normal hierarchy (lower blue) and that for  the inverted hierarchy (upper red) is governed by the size of the matter effects. Notice that the ellipses overlap for a significant fraction of values of the CP--phase $\delta$ for every allowed value of $\sin^2 2
\theta_{13}$.  This makes the determination of sign$(\deltaunotre)$ extremely difficult, i.~e., the sign$(\deltaunotre)$-extraction is not free of degeneracies.

\begin{figure}[t]
\begin{center}
\begin{tabular}{ll}
\includegraphics[width=3in]{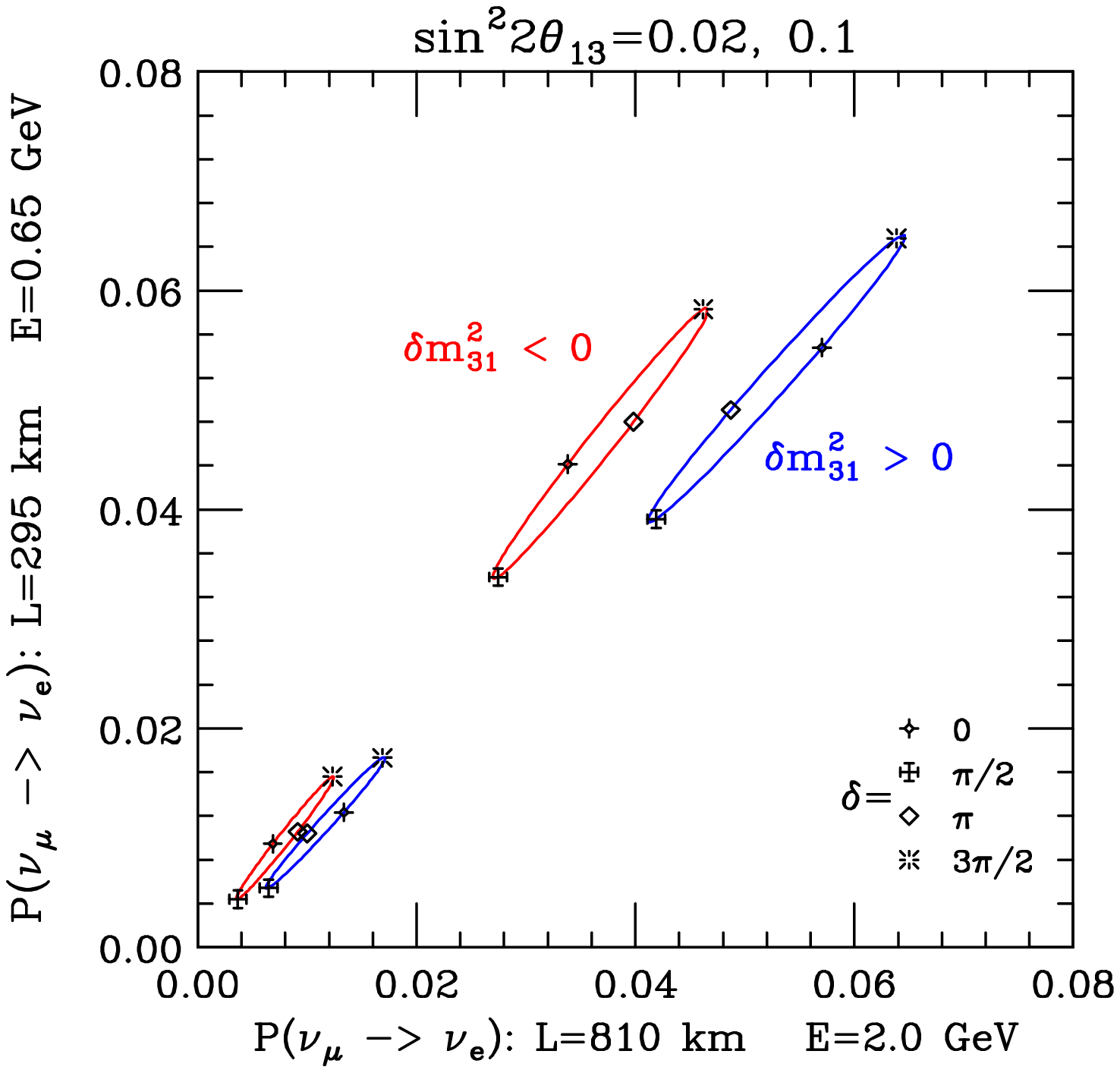}&\hskip 0.cm
\includegraphics[width=3in]{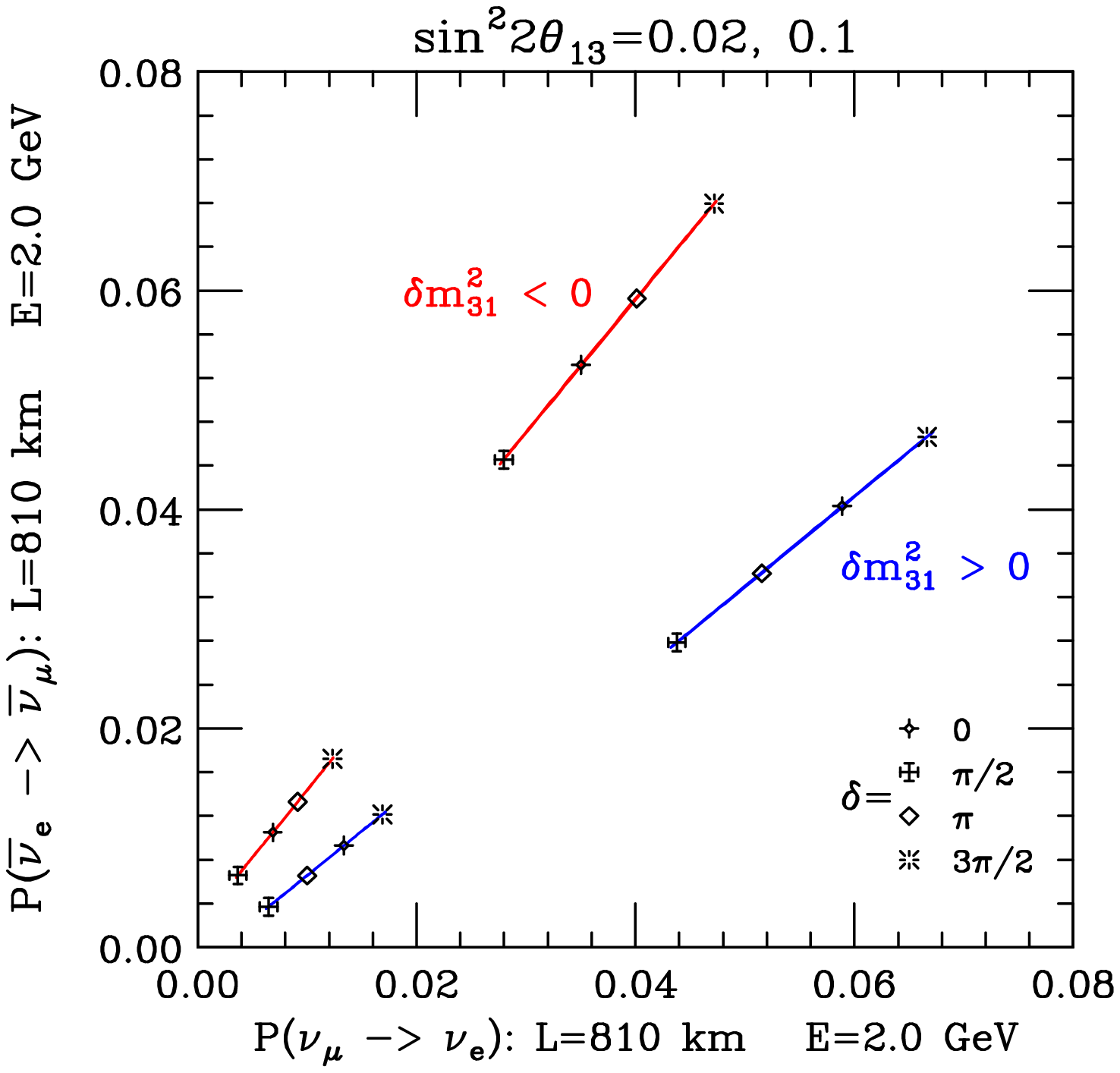}\\
\hskip 2.truecm
{\small (a) Neutrino--Neutrino}            &
\hskip 2.truecm
{\small (b) Neutrino--CPT conjugated channel}
\end{tabular}
\end{center}
\caption[]{\textit{(a) The left panel is the bi--probability plot for $P(\nu_\mu \to \nu_e)$ versus 
$P(\nu_\mu \to \nu_e)$ with baselines 295 km and 810 km for the normal (blue) and the inverted (red) hierarchies.
The smaller, lower (larger, upper) ellipses are for $\sin^2 2 \theta_{13}=0.02$ ($~0.10)$.
The mean neutrino energies are chosen such that the $\langle E \rangle /L$ for the two experiments are approximately identical.\\
(b) The right panel is the bi--probability plot for $P(\nu_\mu \to \nu_e)$ versus 
$P(\bar{\nu}_e \to \bar{\nu}_\mu)$ for the normal (blue) and the inverted (red) hierarchies. 
The baseline and mean neutrino energy for both experiments are 810 km and  $\sim$ 2 GeV, respectively.
The smaller, lower (larger, upper) squashed ellipses are for $\sin^2 2 \theta_{13}=0.02$ ($~0.10)$.
} }
\label{fig:comp2}
\end{figure}

Following the line of thought developed by Minakata, Nunokawa and Parke~\cite{MNP03}, we exploited in a previous work~\cite{mmnp,yo} the neutrino data only from two experiments at different distances and at different off-axis locations, such that the $\langle E \rangle /L$ is the same for the two experiments (see also Refs.~\cite{HLW02,BMW02, SN1,SN2,twodetect}). In the case of bi--probability plots for neutrino--neutrino modes at different distances (which will be referred as near (N) and far (F)), the CP--trajectory is also elliptical. 
In Figure~\ref{fig:comp2}~(a) we present the bi--probability plot for the mean energies and baselines of the $\nu_e$ appearance experiments 
T2K~\cite{T2K} and NO$\nu$A~\cite{newNOvA}. 
The overlap of the two ellipses, which implies the presence of a degeneracy of the type of hierarchy with other
parameters, is determined by their width and the difference in the slopes. Using the fact that matter effects are small ($aL\ll\Delta_{31}$, being $a = G_F N_e/\sqrt{2} \approx (4000~km)^{-1}$ the matter parameter), we can perform a perturbative expansion and assuming that the $\langle E \rangle /L$ of the near and far experiments is the same\footnote{The reason for this choice of $\langle E \rangle /L$ is explained in the next paragraph.}, at first order, 
the ratio of the slopes reads~\cite{MNP03}
\beq
\frac{\alpha_+}{\alpha_-} \simeq
1 +  4 \left( a_{\rm N} L_{\rm N} - a_{\rm F} L_{\rm F} \right)\left( \frac{1}{\Delta_{31}} - \frac{1}{\tan(\Delta_{31})} \right)~,
\label{eq:ratioapp}
\eeq
where $\alpha_+$ and $\alpha_-$ are the slopes of the center of the ellipses as one varies $\theta_{13}$ for normal and inverted
hierarchies, $a_{\rm F}$ and $a_{\rm N}$ are the matter parameters, and $L_{\rm F}$ and $L_{\rm N}$ are the baselines for the two experiments. 
The separation between the center of the ellipses for the two hierarchies increases 
as the difference in the matter parameter times the path length, ($aL$), 
for the two experiments increases. 
Also, since $(\Delta^{-1} - \cot{\Delta})$ is a
monotonically increasing function of $\Delta$, we conclude that the smaller the energy, 
the larger the ratio of slopes, assuming the same $\langle E \rangle /L$.
However the width of the ellipses is crucial: even when the separation 
between the central axes of the two regions is substantial, if the ellipses  for
the normal and inverted hierarchy overlap, the hierarchy cannot be resolved for
values of the CP phase, $\delta$, for which there is overlap.
The width of the ellipses is determined by the difference in the 
$\langle E \rangle /L$ of the two experiments.

In the case of bi--probability plots for the $\nu_\mu \to \nu_e$ and its CPT conjugated channel $\bar{\nu}_e \to \bar{\nu}_\mu$ 
at the same energy divided by baseline,$\langle E \rangle /L$, the CPT--trajectory collapses to a line (see Figure~\ref{fig:comp2}~(b)). 
As for the neutrino-neutrino case, we can perform a
perturbative expansion, and, assuming that the $\langle E \rangle /L$ of the CPT conjugated channels is the same (to minimize the ellipses width), at first order, the ratio of the slopes reads (see Appendix and also Ref.~\cite{MNP03})
\beq
\frac{\alpha_+}{\alpha_-} \simeq
1 +  4 \left( a L + a_{\rm CPT} L_{\rm CPT} \right)\left( \frac{1}{\Delta_{31}} - \frac{1}{\tan(\Delta_{31})} \right)~,
\label{eq:ratiocpt}
\eeq
where $\alpha_+$ and $\alpha_-$ are the slopes of the center of the ellipses as one varies $\theta_{13}$ for normal and 
inverted hierarchies, $a$ and $a_{\rm CPT}$ are the matter parameters and $L$ and $L_{\rm CPT}$ are the baselines for the
 two experiments which exploit the $\nu_\mu \to \nu_e$ and its CPT conjugated channel ($\bar{\nu}_e \to \bar{\nu}_\mu$). 
Notice that, compared to the neutrino--neutrino case given by Eq.~(\ref{eq:ratioapp}), the separation between the center of the ellipses for the two hierarchies increases as the
 sum of the matter parameter times the baseline, $aL$, for both experiments does. Here the ratio of the slopes is enhanced by matter effects for both $\nu_\mu \to \nu_e$ and its CPT conjugated channel $\bar{\nu}_e \to \bar{\nu}_\mu$. Figure~\ref{fig:comp2}~(b) shows the bi--probability curves for the combination of these two flavor transitions, assuming that the two experiments are performed at the same mean energy and baseline. If the $\langle E \rangle /L$ of both experiments is the same, the ellipses will become lines with a negligible width. The separation of the lines for the normal and inverted hierarchy grows as the matter effects for both experiments increase.    
Consequently, the comparison of CPT conjugated channels is more sensitive to the neutrino mass hierarchy than the neutrino--neutrino one.

\section{Beta Beams at Fermilab}
\label{sec:beta}
A betabeam facility exploits a beam of electron neutrinos (antineutrinos) from boosted-ion $\beta^{+}$ ($\beta^{-}$) decays in the straight section of a storage ring~\cite{zucchelli,mauro}. The idea of considering higher $\gamma$ factors (and, consequently, longer detector baselines) was first proposed in Ref.~\cite{betabeampilar}. An extensive phenomenological work has been devoted in order to optimize the betabeam physics reach, analyzing several scenarios with different $\gamma$ factors, boosted-ions and/or detector baselines~\cite{betabeams}.

Early on, $^6$He and $^{18}$Ne were identified as optimal ions, because of the low Q factor of their decay. The lower the neutrino energy is in the rest frame, the more boost is needed to get to a given energy, and since the angular spread of the beam goes as $1/\gamma$ this yields a more focused neutrino beam, which in turn produces more events in the far detector.
Recently, it was proposed to use $^8$Li and $^8$B, which could potentially be produced in large amounts using a small storage ring with an internal gas target\cite{rubbia}. Since these ions have larger Q factor, they produce fewer neutrinos per ion in the far detector for a fixed neutrino energy and baseline. However, because less boost is needed, a smaller accelerator would be needed to achieve the same neutrino energy, as compared to the case of $^6$He and $^{18}$Ne.
In this section we present possible betabeam scenarios at Fermilab, exploiting its current accelerator facilities. Since the analysis considered in this paper only requires anti-neutrinos from a beta-beam, we concentrate on $^6$He and $^8$Li.

A rather thorough study of achievable ion intensities has been done at CERN~\cite{cern-study}. The CERN scenario uses the existing PS and SPS accelerators, and would in addition require a proton source (e.g. the proposed Superconducting Proton Linac), target station, ion source, ion linac and Rapid Cycling Synchrotron (RCS), as well as a decay ring operating at SPS top energy. 
Based on a $^6$He ion production rate of $2 \times 10^{13}$/s, ions decaying in the straight section directed towards the experiment would produce approximately $3 \times 10^{18}$ antineutrinos per year.
We consider two possible scenarios at Fermilab, namely:
1) accelerating $^6$He to Tevatron top energy and 2) accelerating $^8$Li to Main Injector (MI) top energy. These two scenarios produce neutrinos of comparable energies. The ions would be generated using a proton source (e.g. the Project X linac) and accelerated using e.g. a linac and a small RCS before being injected into the existing Booster. Possibly, the Recycler could also be used to accumulate bunches while the MI is ramping.  In both cases, a new decay ring would be needed to store the ions~\footnote{Note that if the Tevatron top energy is used ($^6$He), the decay ring would be very large and expensive.}.

Extrapolating from the work done at CERN, it appears reasonable to expect a useful ion decay rate (decays in the direction of the experiment) of about $1\times 10^{18}$ $^6$He per year in the Tevatron case. At this intensity, the average power deposition in the Tevatron would be about $1$~W/m, which is a generally accepted limit for hands-on-maintenance. Preliminary simulations indicate that the Tevatron magnets would be able to handle the distributed energy deposition from decay products in the arcs, but special care would have to be taken to cope with the build-up of decay products in the straight sections.
In the case of $^8$Li in MI, the injectors could operate with a significantly higher duty factor, since there is no need to wait for the slow Tevatron ramp. However, at repetition rates and intensities corresponding to a useful ion decay rate of about $1 \times 10^{19}$ per year, activation of the Booster from decay products would become a serious issue.

A very important property of the neutrino beam is the duty factor, defined as the relative fraction of time occupied by the neutrino pulse. This is used to suppress background by gating the data acquisition in the experiment. In the CERN study, a duty factor of a few per mil was obtained with considerable difficulty.
A small duty factor is challenging because it requires the ions to be concentrated in very few bunches, which among other things can cause space-charge problems, in particular at low energies. Using the CERN number of $1\times 10^{12}$ ions per RCS cycle, and a Booster injection energy of around $500$~MeV/u, the beam must be distributed over about 10 bunches in the Booster to keep the space charge tune shift at an acceptable level. 
Approximately eight transfers from the booster per MI cycle would be required to obtain $1\times 10^{19}$ useful $^8$Li decays per year at MI top energy.  Without RF manipulations, this would yield a neutrino beam duty factor of about $10\%$, but a duty factor of order $1\%$ could likely be obtained by coalescing bunches at MI top energy\footnote{Although coalescing is standard procedure for generating single  proton bunches from 53 MHz booster beam, the stability of the coalescing process needs to be demonstrated when generating multiple intense coalesced bunches simultaneously.}. 
In the case of $^6$He, about twenty booster injections per cycle would be required per cycle, in order to compensate for the long Tevatron cycle and obtain a useful $^6$He decay rate of $1\times 10^{18}$ per year. Assuming bunches are coalesced in the MI, this should also yield a neutrino beam duty factor of about $1\%$.
Space charge is not expected to be an issue in the MI or Tevatron at these bunch intensities.

For the $^6$He case, therefore, it appears possible to generate $10^{18}$ useful ion decays per year with a maximum gamma of $\gamma_{\rm He} = 350$. For the $^8$Li case, at a maximal gamma of $\gamma_{\rm Li} = 55$, the rate could be higher (as explained above). We will explore an optimistic scenario of $5 \times 10^{19}$ useful ion decays per year, as well a more pessimistic scenario of $10^{19}$ useful ion decays per year from $^8$Li decays. We will assume a duty factor of $1\%$ for both ion species. Table~\ref{tab:tabx} shows the maximum Lorentz gamma factors in the Fermilab machines for the $^6$He and $^8$Li, as well as other ions considered in the literature.

\begin{table}[bht]
\centering
\begin{tabular}{|| c| c | c | c | c ||}
\hline\hline
Machine & $^6$He$^{2+}$ & $^8$Li$^{3+}$ & $^{18}$Ne$^{10+}$ & $^8$B$^{5+}$ \\
\hline\hline
 Linac & 1.6 & 1.07 & 1.15 & 1.19\\
\hline
 Booster & 3.3 & 3.7 & 5.4 & 6.1\\
\hline
 Main Injector & 54 & 60 & 90 & 101\\
\hline
 Tevatron & 351 & 395 & 586 & 659\\
\hline \hline
\end{tabular}
\caption{\it{Maximum Lorentz gamma factors obtainable in the Fermilab machines for different ions of interest.}}
\label{tab:tabx}
\end{table} 

\section{Possible experimental setups at Fermilab}
\label{sec:possible}

As we discussed in Section~\ref{sec:strat}, the most sensitive, degeneracy free method, to extract the neutrino mass hierarchy 
exploiting a future high intensity conventional neutrino beam ($\nu_\mu \to \nu_e$) 
is the combination of this channel with its CPT conjugate $\bar{\nu}_e \to \bar{\nu}_\mu$.
Future facilities like betabeams (neutrino factories) can produce  neutrino beams which are 
entirely (partially) composed of $\nu_e$ or $\bar{\nu}_e$.

The experimental strategy that we follow here is to combine the NO$\nu$A experiment, 
which will measure the flavor transitions $\nu_\mu \to \nu_e$, 
with its CPT conjugated channel. 
The NO$\nu$A experiment is expected to run at least five years with neutrinos.
A $30$ kton low density tracking calorimeter with an efficiency of $24\%$ would 
be located at a baseline of $810$~km and at $12$~km off-axis distance 
from the beam center, resulting in a mean muon neutrino energy of $2.0$ GeV. 
For the CPT conjugate channel, $\bar{\nu}_e\to \bar{\nu}_\mu$, we exploit possible, 
future betabeam facilities at Fermilab 
described in the previous section 
for two antineutrino emitters: $^6$He and $^8$Li. 
A future neutrino factory exploiting neutrino fluxes from muon decays could also provide the $\bar{\nu}_e$ CPT conjugate channel, 
if $\mu^{-}$'s are stored in the decay ring. In the present study we do not explore this possibility.
For the $^6$He ion case, (maximum $\gamma_{\textrm{He}}=350$), the mean electron antineutrino energy is $\sim 1.2$~GeV
($\langle E_\nu \rangle \simeq \gamma_{\textrm{He}} E_{0,\textrm{He}}$, 
$E_{0,\textrm{He}}=3.5 + m_e$~MeV, being the electron end-point energy for $^6$He).
We present the results for two possible experimental $^6$He setups. 
In the first scenario, 
we consider a single baseline $L=810$ km, at which the NO$\nu$A detector would be located, 
with a total detector mass of $40$~kton. 
This first scenario could be 
easily achieved adding to the $30$~kton NO$\nu$A far detector a second $10$~kton detector at 
the NO$\nu$A far site. Possible detector technologies are Liquid Argon or iron calorimeter detectors. 
Liquid Argon (LAr) detectors have excellent efficiency, background rejection and energy resolution, but they could suffer 
from a large atmospheric neutrino background, which could be overcome only if the beam duty cycle 
is $<10^{-4}$~\cite{mauro}. 
In the second scenario, we consider a detector similar to the one of the 
MINOS~\cite{MINOSprop} experiment ($5$~kton) at $735$~km but twice in size. 
If the ion luminosity could be improved by a factor of two, MINOS far detector would 
be sufficient. 
This scenario benefits from the lower atmospheric neutrino backgrounds at the MINOS site. The beam duty cycle 
would not, therefore, be a major issue, and a duty factor of $\sim 1\%$ would be sufficient to overcome the atmospheric 
neutrino background in this case.  
For the $^8$Li ion case, (maximum $\gamma_{\textrm{Li}}=50$), the mean electron antineutrino energy is $\sim 1.8$~GeV
($\langle E_\nu \rangle \simeq \gamma_{\textrm{Li}} E_{0,\textrm{Li}}$, 
$E_{0,\textrm{Li}}=16$~MeV, being the electron end-point energy for $^8$Li).
In order to ensure an almost degeneracy free hierarchy 
measurement, the $\langle E \rangle /L$ of the $\nu_\mu \to \nu_e$ channel 
from NO$\nu$A and its CPT conjugate channel should be similar, 
therefore the detector should be located at $L=300$~km. 
We will consider  $5 \times 10^{19}$ ($1 \times 10^{19}$) 
useful ion decays per year with a  $10$ ($50$)~kton detector respectively. 
As previously discussed, considering $5 \times 10^{19}$ 
useful ion decays is a quite optimistic assumption. 
However, notice that the same statistics could be achieved with 
the more conservative assumption of $1 \times 10^{19}$ useful ion decays 
per year, with a larger $50$~kton detector (these two configurations 
provide the same statistics). 
Due to the lower neutrino energies, 
the detector could be a Liquid Argon TPC, a NO$\nu$A-like Totally Active 
Scintillator Detector (TASD), or a water Cherenkov.
For the NUMI off-axis neutrino beam, we have not considered a binning in the signal, 
since neutrino events will lie in a very narrow energy window [1.5,2.5] GeV.
The $^6$He and $^8$Li antineutrino betabeams have been divided in 
three energy bins, assuming an energy detection threshold of $0.5$~GeV. 
For the $^6$He case, we divide the signal in three energy bins, 
in the energy ranges [0.5,1.0], [1.0,1.5] and [1.5,2.4] GeV, 
respectively. In the $^8$Li case, the energy ranges are [0.5,0.75], [0.75,1.0] and [1.0,1.5] GeV, respectively. 

\begin{figure}[t]
\begin{center}
\begin{tabular}{ll}
\includegraphics[width=3in]{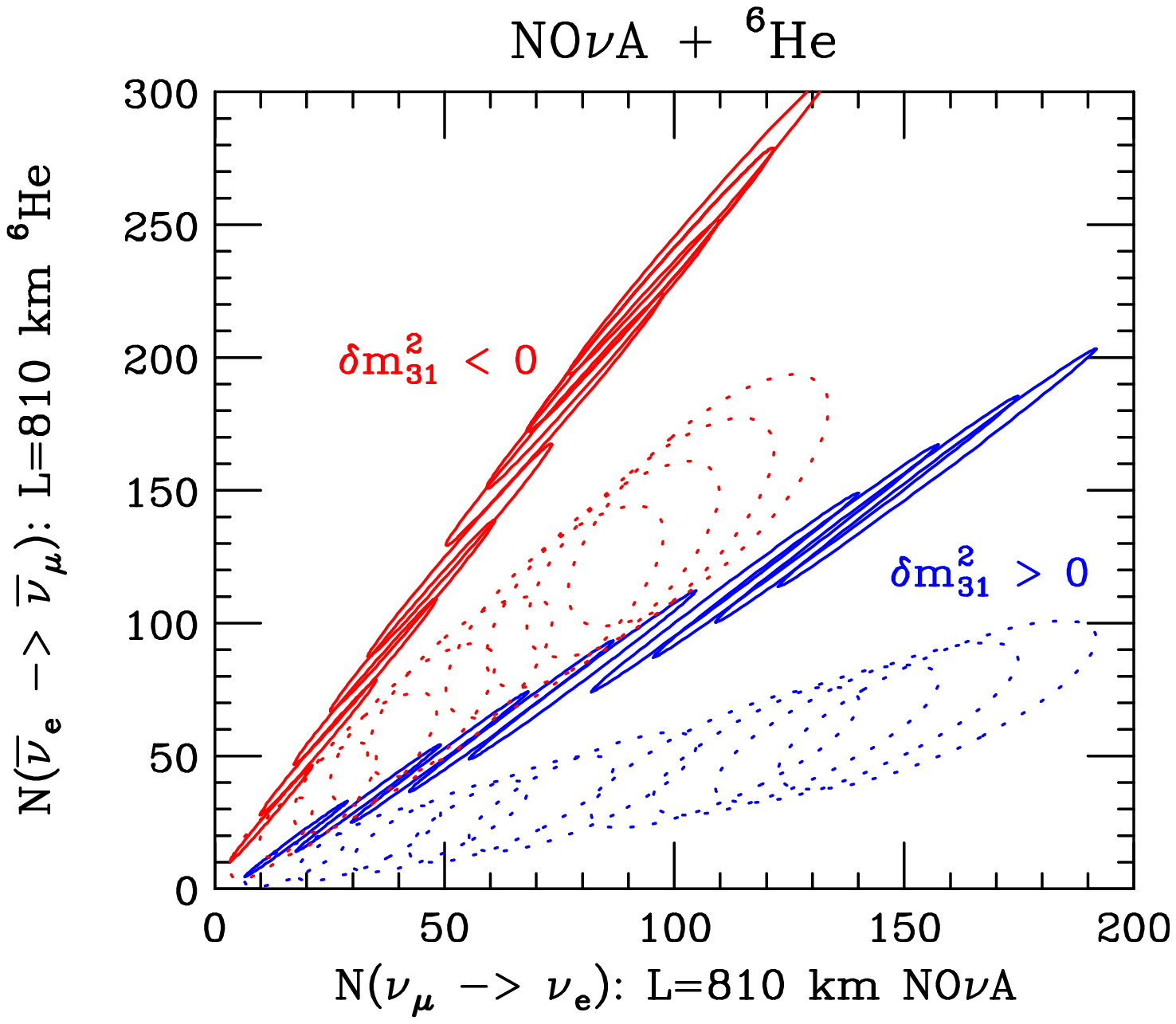}&\hskip 0.cm
\includegraphics[width=3in]{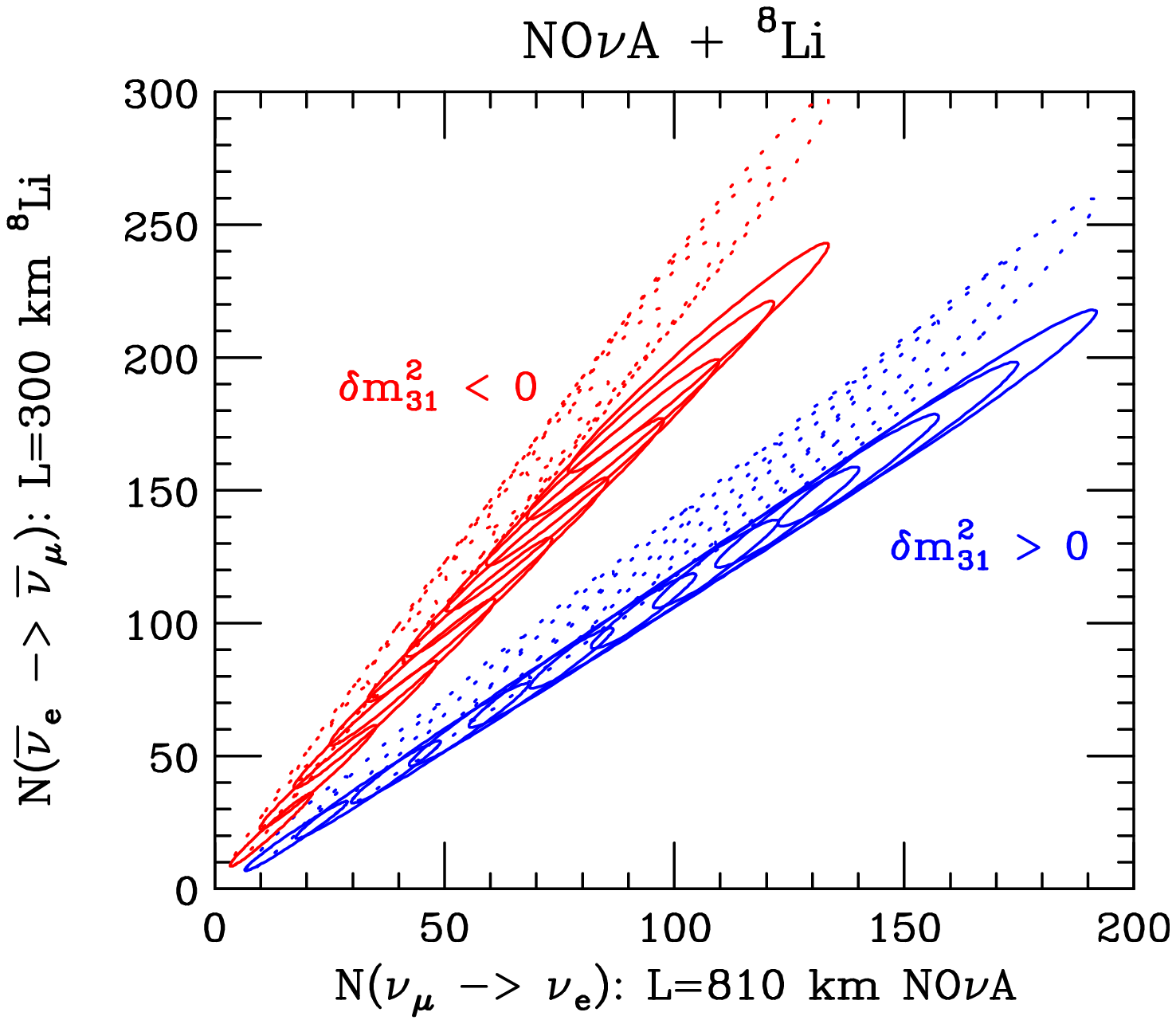}\\
\hskip 2.truecm
{\small (a) NO$\nu$A--$^6$He $\beta$eta-beam} &
\hskip 2.truecm
{\small (b) NO$\nu$A--$^8$Li $\beta$eta-beam}
\end{tabular}
\end{center}
\caption[]{\textit{(a) The allowed regions in the bi--event plot for 
$N(\nu_\mu \to \nu_e)$ for NO$\nu$A versus $N(\bar{\nu}_e \to 
\bar{\nu}_\mu)$ for a betabeam experiment which exploits antineutrinos 
from $^6$He decays with $\gamma_{He}=350$, and a detector of $40$~kton 
located at a distance of $L=810$~km. The blue (red) ellipses denote normal (inverted) hierarchies. From bottom up, 
the ellipses correspond to $\sin^2 2 \theta_{13}$ varying from $0.01$ to $0.1$. The solid (dashed) ellipses illustrate the third (second) energy bins of the betabeam spectrum. (b) Same as (a) but with 
antineutrino fluxes resulting from  $^8$Li decays, and a detector of $10$~kton at $300$~km.}}
\label{fig:bievents}
\end{figure}

Figure~\ref{fig:bievents} depict the bi--event plots for the combination of the NO$\nu$A neutrino events 
($\nu_\mu \to \nu_e$) with its CPT conjugated channel ($\bar{\nu}_e\to\bar{\nu}_\mu$) 
from the betabeam experiment, resulting from the decays of $^6$He (left panel) and $^8$Li 
(right panel), for both normal and inverted hierarchies. The statistics considered for NO$\nu$A 
correspond to Phase I of the experiment (neutrino running only). 
For the $^6$He betabeam experiment, we assume a number of useful ion decays  of $10^{18}$ per 
year, five years of data taking, and a $40$~kton detector located at $810$~km. 
For the $^8$Li betabeam experiment, we assume $5 \times 10^{19}$ ($1 \times 10^{19}$) useful ion decays per year, 
ten years of data taking, and a $10$ ($50$)~kton detector located at $300$~km. 

Figure~\ref{fig:bievents}~(a) shows that, for the combination of NO$\nu$A off-axis neutrino
events  with the antineutrino events from $^6$He decays, the separation
between the bi--event contours for the normal and inverted hierarchies 
is larger than in the case of  $^8$Li generated antineutrino
events, as seen in Figure~\ref{fig:bievents}~(b). 
As previously explained, 
the difference in the slopes of the two hierarchies is proportional to
the sum of the size of matter effects times the baseline, 
$a_{NO\nu A} L_{NO\nu A} + a_{CPT} L_{CPT}$. The product $a_{CPT}L_{CPT}$ is
larger for the $^6$He betabeam $\bar{\nu}_e$ events (with a baseline of $810$~km), 
than for the $^8$Li betabeam $\bar{\nu}_e$ events (with
a baseline of $300$~km). 
The solid (dashed) contours in Figure~\ref{fig:bievents} \
show the number of betabeam antineutrino events in the second (third) energy bin. 
When the $\langle E \rangle /L$ 
of the $\nu_\mu \to \nu_e$ and its CPT conjugated channel are similar, 
the ellipses width is minimal (they collapse to a line) and therefore 
the elliptical contours for the normal and inverted hierarchies 
will not overlap, regardless the value of the CP violating phase $\delta$. 
For the combination of NO$\nu$A off-axis neutrino
events with the $^6$He betabeam antineutrino events, 
there exists a clear difference between the second and third 
energy bins in the bi--event contours: while they are ellipses in the former, 
they are almost lines in the latter. Only in the third energy bin 
([1.5,2.5] GeV) is the $\langle E \rangle /L$ the same for the $^6$He 
betabeam $\bar{\nu}_e$ and for the NO$\nu$A $\nu_\mu$ events.  
For the $^8$Li case the ellipses width is minimal for both 
the second and third energy bins, since both bins are close to $E\sim 0.8$~GeV, 
the energy at which the $(\langle E \rangle /L)_{Li}$ equals the  $(\langle E \rangle /L)_{NO\nu A}$.

\section{Sign $\Delta m^2_{31}$ sensitivities}
\label{sec:results}
In this section we present the physics results from the combination of antineutrino data, resulting from
several possible betabeam setups, with the NO$\nu$A neutrino data.

\begin{figure}[h]
\begin{center}
\begin{tabular}{ll}
\includegraphics[width=3in]{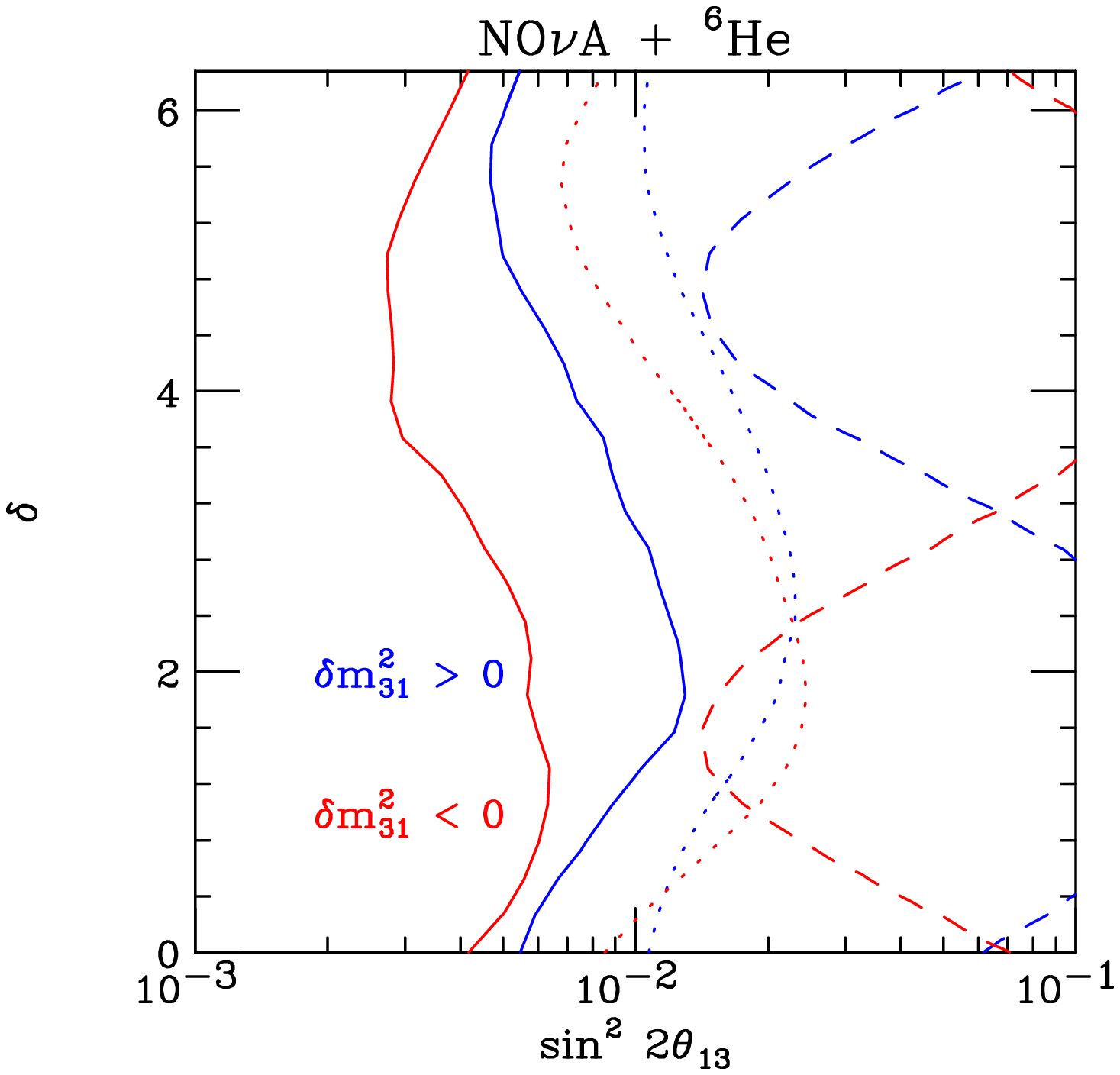}&\hskip 0.cm
\includegraphics[width=3in]{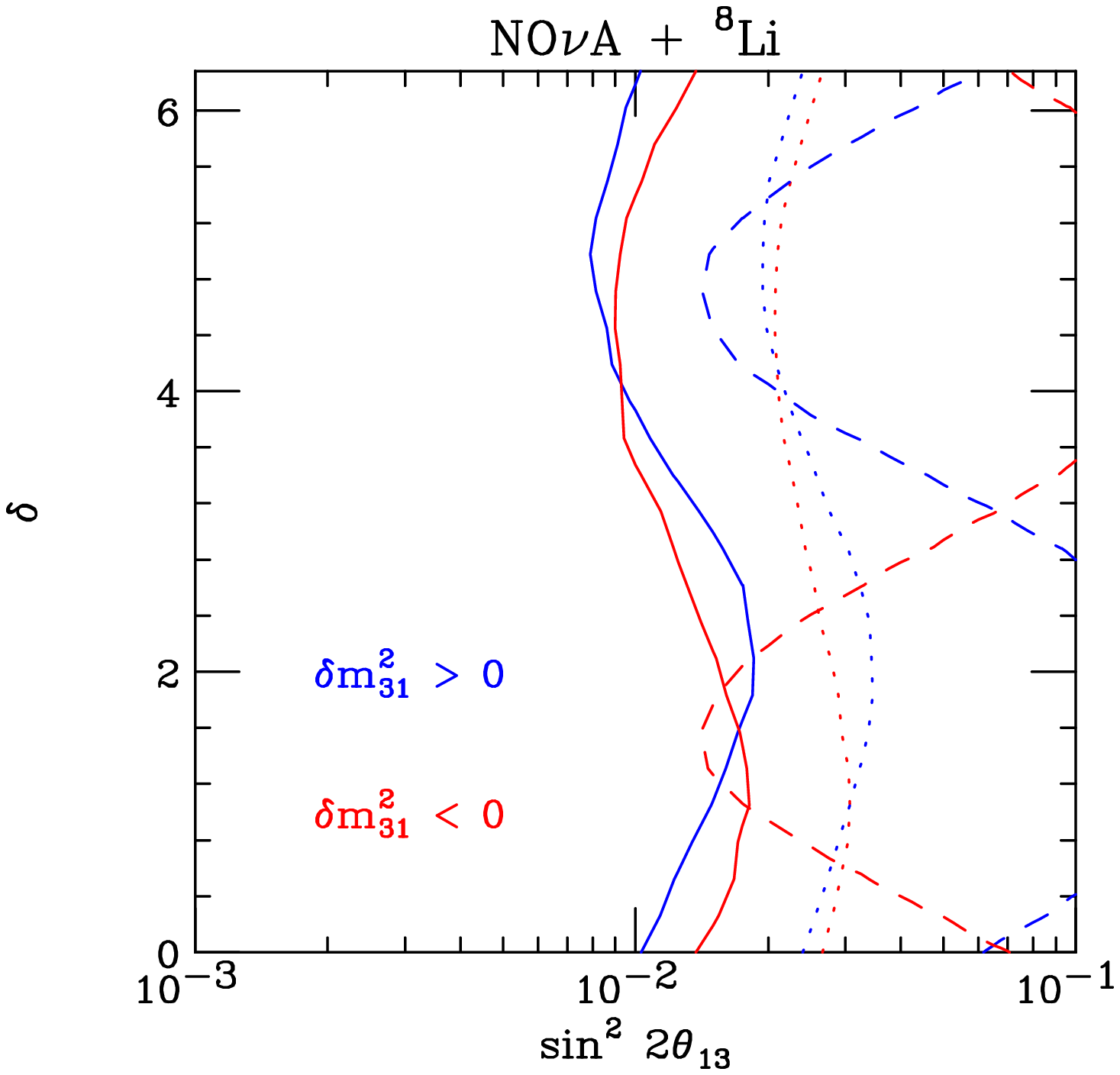}\\
\hskip 4.0truecm
{\small (a)} &
\hskip 4.0truecm
{\small (b)}
\end{tabular}
\end{center}
\caption[]{\textit{(a) The $90\%$ CL (2 d.o.f) hierarchy resolution curves for different exposures for the $^6$He betabeam  $\bar{\nu}_e$ fluxes at $810$~km, combined with five years of neutrino data only from the NO$\nu$A far detector, located $12$~km off-axis at $810$~km. Only backgrounds in the NO$\nu$A experiment have been included. The blue (red) curves assume normal (inverted) hierarchy. The solid (dotted) line depicts the results for $2 \times 10^{20}$ ($5\times 10^{19}$) useful ion decays times  kton. The blue (red) dashed curve shows the sensitivity reach at $90\%$ CL (2 d.o.f.) from the combination of neutrino and antineutrino data from the NO$\nu$A experiment, assuming five years running in both neutrinos and antineutrinos with a factor of five increase in statistics and normal (inverted) hierarchy. 
(b) The $90\%$ CL (2 d.o.f) hierarchy resolution curves for different exposures for the $^8Li$ betabeam  electron antineutrino fluxes at $300$~km, combined with muon neutrino data only from the NO$\nu$A far detector at $12$~km off-axis at $810$~km. Again, only backgrounds in the NO$\nu$A experiment have been included.The blue (red) curves assume normal (inverted) hierarchy. The solid (dotted) line depicts the results for $5 \times 10^{21}$ ($10^{21}$) useful ion decays times kton.}}
\label{fig:comp1}
\end{figure}

Figure~\ref{fig:comp1}~(a) shows the $90\%$ C.L mass hierarchy sensitivity, assuming 
two degrees of freedom statistics (2 d.o.f, that is, $\Delta \chi^2 > 4.21$)
for the combination of NO$\nu$A neutrino data with the $^6$He betabeam antineutrino data, 
neglecting the background in the $\bar{\nu}_e \to \bar{\nu}_\mu$ channel. 
In order to safely neglect the atmospheric neutrino background, one would 
need a very small, experimentally challenging duty factor for the betabeam neutrino fluxes. 
The baseline for the two experiments is fixed at $810$~km and the $\gamma_{\textrm{He}}=350$, 
in order to have a similar $\langle E \rangle/L$ in both the muon neutrino and the electron antineutrino channels. 
The binning of the signal has been chosen as quoted in the previous section. We have considered a 
flux of $10^{18}$ antineutrinos per year, and five years of data taking. The blue (red) 
lines assume normal (inverted) hierarchy, and the solid (dotted) lines depict the results 
for a betabeam antineutrino experiment with a 40 (10)~kton detector. For the 
$40$~kton detector, the sensitivity is better for the inverted mass hierarchy, since 
in this case statistics is dominated by the antineutrino channel $\bar{\nu}_e \to \bar{\nu}_\mu$. 
For the $10$~kton detector, the two channels (i.e., the neutrino channel 
from NO$\nu$A and the antineutrino channel from the betabeam experiment) will have similar statistics 
and therefore the sensitivity for the normal and the inverted mass hierarchies are similar. 
As a comparison, we also show the results for the NO$\nu$A experiment (upgraded by a factor of five in statistics), 
assuming five years of neutrino and five years of antineutrino data taking, see the dashed curves in Figure~\ref{fig:comp1}. 
The setup proposed here improves the sensitivity of the NO$\nu$A upgraded experiment by an order of magnitude, 
and more importantly, eliminates the dependence of the mass hierarchy determination on the value 
of the CP violating phase $\delta$.

Figure~\ref{fig:comp1}~(b) shows the $90\%$ C.L mass hierarchy sensitivity, 
assuming two degrees of freedom statistics (2 d.o.f, that is, $\Delta \chi^2 > 4.21$)
for the combination of NO$\nu$A neutrino data with the $^6$Li betabeam antineutrino data, neglecting 
the background in the $\bar{\nu}_e \to \bar{\nu}_\mu$ channel. The baseline for the 
betabeam experiment is $L=300$~km and the binning of the signal has been chosen as 
described in the previous section. 
We assume $5 \times 10^{19}$ $^8$Li generated antineutrinos per year, and ten years of
data taking. The blue (red) lines assume normal (inverted) hierarchy, and the 
the solid (dotted) lines depict the results 
for a betabeam antineutrino experiment with a 10 (2)~kton detector. For the case of $^8$Li betabeam experiment, the sensitivity is similar 
for the normal and inverted mass hierarchy, and smaller than in the 
case of $^6$He betabeam experiment. This is expected, since at a shorter baseline ($300$~km) 
the product of the matter potential times the distance is reduced. Again, the combination of 
the NO$\nu$A neutrino data only with the $^8$Li betabeam antineutrino data provides a much better 
sensitivity to the mass hierarchy than the NO$\nu$A upgraded experiment alone.

\begin{figure}[h]
\begin{center}
\begin{tabular}{ll}
\includegraphics[width=3in]{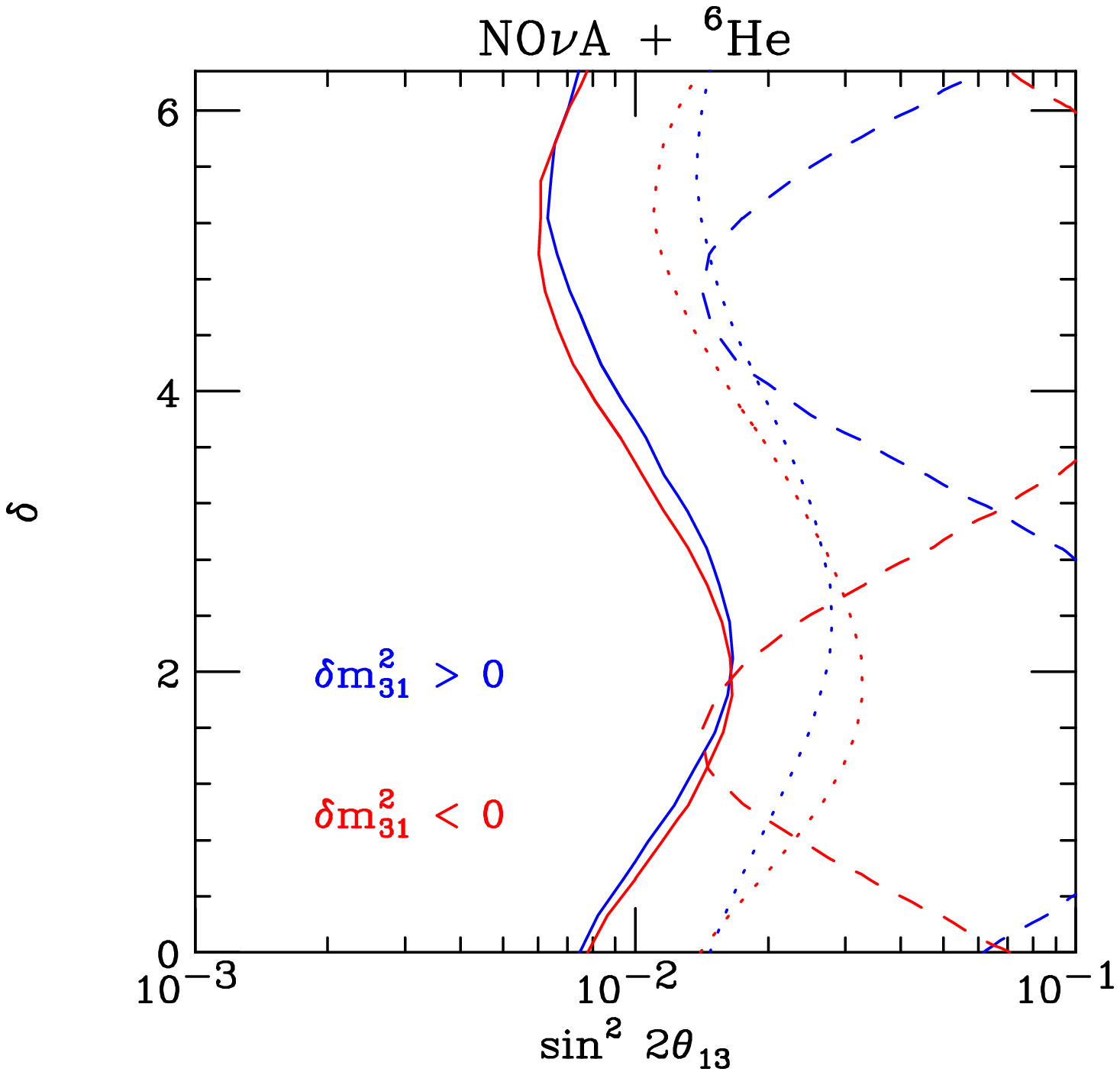}&\hskip 0.cm
\includegraphics[width=3in]{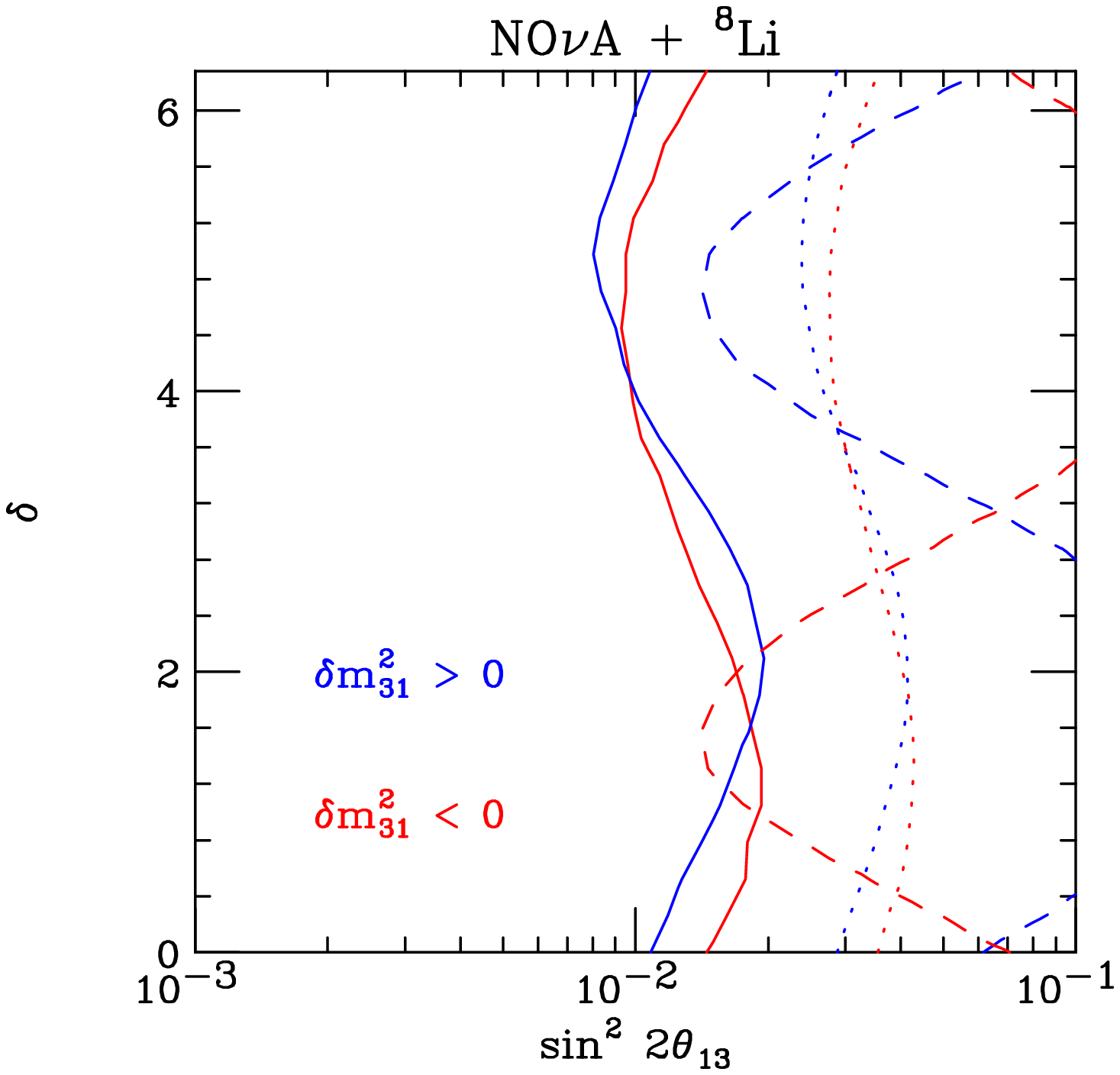}\\
\hskip 4.0truecm
{\small (a)} &
\hskip 4.0truecm
{\small (b)}
\end{tabular}
\end{center}
\caption[]{\textit{ Same as Figs.~\ref{fig:comp1} but including backgrounds in the betabeam electron antineutrino data, see text for details.}}
\label{fig:compx}
\end{figure}

However, as previously stated, the beam duty cycle needed in order to neglect 
the atmospheric neutrino background is highly challenging. For a MINOS-like detector, 
there are $30$ atmospheric neutrino interactions per kton-year which could mimic a muon 
coming from the oscillated $\bar{\nu}_e \to \bar{\nu}_\mu$~\cite{andy}. In order to avoid 
such a large background, we have assumed a betabeam duty cycle $\sim10^{-2}$ , 
which seems experimentally achievable. Figure~\ref{fig:compx} shows the 
equivalent to Figure~\ref{fig:comp1} when adding the atmospheric neutrino background quoted above, 
rescaled accordingly to the detector sizes and the exposure times. 
Notice that the presence of a non negligible atmospheric neutrino background in the 
antineutrino channel $\bar{\nu}_e \to \bar{\nu}_\mu$ reduces the sensitivity reach 
especially for the case of the inverted mass hierarchy. However, 
even in the presence of a non negligible background, and with the very conservative 
assumption of a $10^{-2}$ beam duty cycle, the combination of the betabeam antineutrino data 
with the NO$\nu$A Phase I neutrino data, provides better sensitivity than the 
NO$\nu$A experiment alone (upgraded by a factor of five and running in both neutrino and antineutrino mode). 
If a smaller duty factor $<10^{-2}$ could be achievable (as commonly assumed, following Ref.~\cite{mauro}) the sensitivity to the mass hierarchy 
would lie within the limits illustrated in Figure~\ref{fig:comp1} 
(the most optimistic case with no atmospheric neutrino induced background) 
and the limits depicted in Figure~\ref{fig:compx} (the most pessimistic case with atmpospheric neutrino backgrounds and a beam duty cycle $\sim 10^{-2}$). 

\section{Conclusions}
\label{sec:conclusions}
We have explored an alternative strategy for measuring the neutrino mass hierarchy. 
Unlike the approach followed by future long baseline neutrino oscillation
experiments that combine the neutrino--antineutrino data, 
the combination of the CPT conjugated channels that we study here 
provides an almost degeneracy free determination of the neutrino mass hierarchy, 
provided the two channels have similar $\langle E \rangle /L$. 
Future neutrino facilities at Fermilab could provide these CPT neutrino--conjugated channels. 
The NO$\nu$A $\nu_e$ off-axis appearance experiment could provide the $\nu_\mu\to \nu_e$ channel. 
A future betabeam facility based at Fermilab could provide the CPT-conjugated $\bar{\nu}_e \to \bar\nu_\mu$ channel. 
A realistic estimate of the expected electron antineutrino fluxes from boosted ion decays is presented.  
We propose two possible accelerator scenarios for generating the betabeam electron antineutrino fluxes: 
the Tevatron, which could accelerate $^6$He ions, and the Main Injector, which could accelerate $^8$Li ions. In the case of the Tevatron, the decay ring would be very large and possibly prohibitively expensive. 
The first scenario could benefit from the NO$\nu$A far detector at $L=810$~km, (but the decay ring needed would be very large) ; for the second scenario, 
an additional, although smaller $2-10$~kton MINOS like detector at a 
shorter baseline, $L=300$~km, would be necessary (the decay ring needed in this case would be smaller, though). 
In the more pessimistic case, with a modest beam duty cycle of $10^{-2}$ and including realistic atmospheric 
neutrino backgrounds, the neutrino mass hierarchy could be determined for $\sin^2 2 \theta_{13}>0.01$, independently of 
the value of the CP violating phase $\delta$, for both accelerator possibilities. 
These two alternative choices could improved by an order of magnitude the sensitivity to the neutrino mass hierarchy 
obtained by a future NO$\nu$A upgraded experiment exploiting both neutrinos and antineutrinos.

\section*{Acknowledgments}
We wish to thank A.~Donini for useful comments on the manuscript. OM is supported by the European Programme ``The Quest for
Unification''  contract MRTN-CT-2004-503369. Fermilab is operated by FRA under DOE contract DE-AC02-07CH11359. OM would like to thank the Theoretical Physics Department at Fermilab for hospitality and support. 

\appendix
\section{CPT Conjugate Probabilities:}
\label{sec:cptintro}
The amplitudes for $\nu_\mu \to \nu_e$ and $\bar{\nu}_e \to \bar{\nu}_\mu$ consists of two terms,
one associated with the atmospheric $\delta m^2$ scale and the other associated with the solar
$\delta m^2$ scale. 
Thus, the probability for these the CPT-conjugate processes contain three terms; 
the square of each of the amplitudes plus the interference term between the two amplitudes which depends on the CP phase $\delta$.
For the normal (upper sign) and inverted (lower sign) hierarchy, the $\nu_\mu \to \nu_e$ and  $\bar{\nu}_e \to \bar{\nu}_\mu$ appearance probabilities are  given by
\begin{eqnarray}
P(\nu_\mu \to \nu_e) & = & X_\pm \theta^2 \pm 2\sqrt{X_\pm} \sqrt{P_\odot} ~\theta \cos(\pm\Delta_{31}+\delta) + P_\odot \nonumber \\
\overline{P} (\bar{\nu}_e \to \bar{\nu}_\mu) & = & X_\mp  \theta^2 \pm2\sqrt{X_\mp} \sqrt{P_\odot} ~ \theta  \cos(\pm\Delta_{31}+\delta)
+P_\odot .
\label{eq:cptpair}
\end{eqnarray}
The coefficients $P_\odot$ and  $X_{\pm}$ are simply
\begin{eqnarray}
\sqrt{P_\odot} & = & \cos \theta_{23} \sin 2 \theta_{12} \frac{\sin (aL)}{(aL)} ~\Delta_{21} ,
\nonumber \\
\sqrt{X_\pm} & =&  2 \sin \theta_{23} \frac{\sin (\pm \Delta_{31} -aL)}{ (\pm \Delta_{31} -aL)} ~\Delta_{31} ,
 \nonumber 
\end{eqnarray}
where $\Delta_{ij} =|\delta m^2_{ij} | L/4E$ and $a = G_F N_e/\sqrt{2} \approx (4000~km)^{-1}$.  The atmospheric amplitude for $\nu_\mu \to \nu_e$ is $\pm \sqrt{X_\pm} \theta$ whereas the solar amplitude is $\sqrt{P_\odot}$ and the relative phase between these two amplitudes\footnote{The full amplitude for $\nu_\mu \to \nu_e$ is 
$(\pm \sqrt{X_\pm} \theta~e^{-i (\pm\Delta_{31}+\delta)} + \sqrt{P_\odot})$.}
 is $(\pm\Delta_{31}+\delta)$.  In vacuum, $X_+=X_- \equiv X_0$ and the two probabilities are identical, as they must since they are CPT conjugates.
 
 The other related CPT conjugate pair of appearance probabilities, $P(\nu_e \to \nu_\mu)$  and  
 $\overline{P} (\bar{\nu}_\mu \to \bar{\nu}_e)$, can be obtained from the above by changing 
 the sign of $\delta$, as follows
 \begin{eqnarray}
P(\nu_e \to \nu_\mu) & = & X_\pm \theta^2 \pm 2\sqrt{X_\pm} \sqrt{P_\odot} ~\theta \cos(\pm\Delta_{31}-\delta) + P_\odot \nonumber \\
\overline{P} (\bar{\nu}_e \to \bar{\nu}_\mu) & = & X_\mp  \theta^2 \pm2\sqrt{X_\mp} \sqrt{P_\odot} ~ \theta  \cos(\pm\Delta_{31}-\delta)
+P_\odot .
\label{eq:othercptpair}
\end{eqnarray}

 The difference between the first two CPT conjugate appearance probabilities, is given by
 \begin{eqnarray}
 P(\nu_\mu \to \nu_e) - \overline{P} (\bar{\nu}_e \to \bar{\nu}_\mu) & = &
\pm  \theta ~(\sqrt{X_+} - \sqrt{X_-})~\left[(\sqrt{X_+} + \sqrt{X_-})\theta \pm 2\sqrt{P_\odot} \cos (\pm \Delta_{13}+\delta) \right].  \nonumber
 \end{eqnarray}
 This quantity is positive for the normal hierarchy (NH) and negative for the inverted hierarchy (IH), if
 \begin{eqnarray}
 \sqrt{X_+}  & > &\sqrt{X_-} \quad  {\rm and } \quad  \theta   >  2 \sqrt{P_\odot}/(\sqrt{X_+}+\sqrt{X_-}) \approx \sqrt{P_\odot}/\sqrt{X_0} , 
 \end{eqnarray}
 for all values of the CP phase $\delta$.  
 The constraint on $\theta$ requires\footnote{This is the value of $\theta$  at which the atmospheric and solar amplitudes have the same magnitude in vacuum.}
 \begin{eqnarray}
\sin^2 2\theta_{13} >  \frac{ \sin^2 2 \theta_{12}  \Delta^2_{21}}{ \tan^2 \theta_{23} \sin^2 \Delta_{31} }
 \sim 0.001- 0.002,
 \label{eq:equalamps}
\end{eqnarray}
whereas the constraint, $\sqrt{X_+}>\sqrt{X_-}$, is satisfied near the first oscillation maximum provided
$(aL) \ll 1$, i.e. $L \ll 4000~km $.

With these rather weak constraints then 
\begin{eqnarray}
P(\nu_\mu \to \nu_e) & > & \overline{P} (\bar{\nu}_e \to \bar{\nu}_\mu) \quad {\rm for ~NH}
\label{eq:nh} \\
{\rm and} \quad 
P(\nu_\mu \to \nu_e) & < & \overline{P}(\bar{\nu}_e \to \bar{\nu}_\mu)  \quad {\rm for ~IH}
\label{eq:ih} 
\end{eqnarray}
for all values of the CP phase $\delta$.  For the normal (inverted) hierarchy, 
the matter effect enhances  (suppresses) the
$P(\nu_\mu \to \nu_e)$ channel and suppresses (enhances) the  $\overline{P}(\bar{\nu}_e \to \bar{\nu}_\mu) $
channel, thus the matter effect in a sense is used twice.
Of course, the difference between these two appearance probabilities is larger at larger values of
$\theta$ and at larger values of the matter effect.  This is the effect that is exploited here to
determine the neutrino mass hierarchy.

In the $P(\nu_\mu \to \nu_e)$ versus $\overline{P}(\bar{\nu}_e \to \bar{\nu}_\mu) $ plane
the trajectory for fixed value of $\theta$ as the CP phase $\delta$ is varied from
0 to $2\pi$ is in general an ellipse which collapses to a line if the $E/L$ of both channels is
the same.  The centre of this ellipse is given by
\begin{eqnarray}
( \overline{P}(\bar{\nu}_e \to \bar{\nu}_\mu), P(\nu_\mu \to \nu_e)) =(X_\mp \theta^2 + P_\odot, ~X_\pm \theta^2 + P_\odot).
\end{eqnarray}
Thus, as $\theta$ is varied, the centre of the ellipses form lines with slope given by
\begin{eqnarray}
\alpha_+ & \equiv & \frac{X_-}{X_+} \quad {\rm for ~NH} \nonumber \\
{\rm and} \quad  \alpha_- &\equiv & \frac{X_+}{X_-} \quad {\rm for ~IH}.
\end{eqnarray}
If the matter effect is small, $(aL) \ll \Delta_{31}$, one can perform a Taylor series about the vacuum
such that
\begin{eqnarray}
\alpha_\pm = 1 \mp 4(aL)[\Delta_{31}^{-1} - \cot \Delta_{31}] +{\cal O}(aL)^2.
\end{eqnarray}
It is the difference in the slopes of the two lines (for the normal hierarchy, $\alpha_+$ and  for the inverted hierarchy, $\alpha_-$)
which provides the separation between the allowed regions for two hierarchies in the   $P(\nu_\mu \to \nu_e)$ versus $\overline{P}(\bar{\nu}_e \to \bar{\nu}_\mu) $ plane.


\end{document}